\documentstyle[prl,aps,preprint,epsf]{revtex}

\def\gsim{\stackrel{>}{{}_\sim}}
\def\lsim{\stackrel{<}{{}_\sim}}

\begin{document}
\preprint{TU-550, RCNS-98-08, ICRR-Rep-425-98-21}
 
\title{Hadronic Axion Model in Gauge-Mediated Supersymmetry Breaking}
\author{T. Asaka}
\address{Institute for Cosmic Ray Research, University of Tokyo,
  Tanashi 188-8502, Japan}
\author{Masahiro Yamaguchi}
\address{Department of Physics, 
  Tohoku University, Sendai 980-8578, Japan}
\date{May 1998}

\maketitle

\begin{abstract}
A simple hadronic axion model is proposed 
in the framework of gauge-mediated supersymmetry breaking.
Dynamics of Peccei-Quinn symmetry breaking 
is governed by supersymmetry breaking effects
and the Peccei-Quinn breaking scale $f_{PQ}$ is inversely proportional to 
the gravitino mass.
The gravitino mass range which corresponds to the axion window $f_{PQ} \simeq
10^{9}$ GeV -- $10^{13}$ GeV lies in the region 
predicted by gauge-mediated supersymmetry breaking models.
The model is also shown to be cosmologically viable. 
\end{abstract}

\clearpage

The Peccei-Quinn (PQ) mechanism \cite{PecceiQuinn} is so far the most
attractive framework to solve the strong CP problem. The essential
ingredient of this mechanism is a global U(1$)_{PQ}$ symmetry which
is, apart from breaking by the QCD anomaly, spontaneously broken. The
resulting Goldstone mode (axion) acquires a small mass due to the QCD
anomaly.  When one considers the PQ mechanism in the context of
supersymmetry (SUSY), there appears a non-compact flat direction
associated with the U(1)$_{PQ}$ Goldstone mode. The degeneracy of the
vacua is resolved by effects of SUSY breakdown.  The properties of the
PQ mechanism should therefore depend on details of the SUSY breakdown.
In this paper, we would like to propose a simple model which works as
the PQ mechanism in the framework of gauge-mediated SUSY breaking,
while most of the previous study has been done in gravity-mediated
SUSY breaking.

%

%
%
We shall first describe our proposal.
We consider a KSVZ axion model\cite{KSVZ} 
(so called a hadronic axion model).
A gauge-singlet Peccei-Quinn (PQ) multiplet $X$ 
and also new PQ quarks $Q_P$ and $\overline{Q}_P$
({\bf 3} and ${\bf 3}^\ast$ in SU(3)$_C$) are introduced.
Their U(1)$_{PQ}$ charges are assigned as
${\cal Q}[X] = + 1$, ${\cal Q}[Q_P] = -1/2$ and 
${\cal Q}[\overline{Q}_P] = - 1/2$.
The  superpotential of the PQ sector takes
the following simple form:
\begin{eqnarray}
    W_{PQ} = \lambda_P X Q_P \overline{Q}_P,
    \label{sp-PQ}
\end{eqnarray}
where $\lambda_P$ is a coupling constant. Note that no mass parameter is
introduced in the superpotential.
Eq.~(\ref{sp-PQ}) gives the potential of the scalar fields 
in the SUSY limit as
\begin{eqnarray}
    V(X, \widetilde{Q_1}, \widetilde{Q_2})
    &=&
    \lambda_P^2 \left[
        \left| \widetilde{Q_2}^\dagger \widetilde{Q_1} \right|^2
        +
        \left| X \right|^2 
        \left( \left| \widetilde{Q_1} \right|^2 
            +  \left| \widetilde{Q_2} \right|^2  \right)
   \right]
\nonumber \\
    &+& \frac{ g_s^2 }{ 2 } 
        \left[ \widetilde{Q_1}^\dagger T^a \widetilde{Q_1}
           - \widetilde{Q_2}^\dagger T^a \widetilde{Q_2}
        \right]^2,
    \label{v-cl}
\end{eqnarray}
where $g_s$ and $T^a$ denote the QCD coupling and generators.
$X$ is the PQ scalar field, 
and $\widetilde{Q_1}$ and $\widetilde{Q_2}$ are 
PQ squarks.%
\footnote{Here $\widetilde{Q_1} \equiv \widetilde{Q_P}$ and
$\widetilde{Q_2} \equiv \widetilde{\overline{Q}}_P^c$.}

{}From the potential (\ref{v-cl}) one finds a flat direction along the
$X$ axis with $\widetilde{Q_1}=\widetilde{Q_2}=0$.  This comes from
the fact that the superpotential (\ref{sp-PQ}) has an extended
U(1$)_{PQ}$ symmetry \cite{KOY}.  Because it is a holomorphic function, the
superpotential which possesses the conventional U(1$)_{PQ}$ symmetry
is also invariant under the complex form of the U(1) transformation
(i.e. the dilatational transformation).  This extended symmetry
results in the flat direction of the scalar potential (\ref{v-cl})
when the SUSY is an  exact symmetry.  However, the flat direction is
lifted through SUSY breaking effects.

In gauge-mediated SUSY breaking theories (for a review see ref. \cite{GMSB}), 
the standard model (SM) gauge
interaction transmits SUSY breaking effects from a messenger sector to
ordinary squarks and sleptons at two-loop level.  At the same level
the PQ squarks $\widetilde{Q_1}$ and $\widetilde{Q_2}$ also feel SUSY
breaking through the QCD interaction, whereas the PQ scalar $X$ 
is a gauge singlet and thus feels SUSY breaking through the Yukawa interaction (\ref{sp-PQ}) only after the PQ squarks  receive SUSY breaking masses.
Then the induced potential for $X$ is so suppressed
that we should not neglect the SUSY breaking effect mediated by
gravity. Indeed as we will see shortly, the balance between the
gravity-mediated effect and the gauge-mediated effect determines the
minimum of the $X$ field. In the following, we estimate the potential
of the flat direction induced by the SUSY breaking effects
by considering the both mediation mechanisms.

Effects of  SUSY breaking communicated by the gravity 
are expected to induce the soft SUSY breaking mass
to the PQ scalar field comparable to the gravitino mass $m_{3/2}$
which is much smaller than the electroweak scale 
in the gauge-mediated SUSY breaking models
($m_{3/2}\sim 10^{-2}$ keV--1 GeV).
Then through the  gravity-mediation the flat direction
obtains the potential $V_H$ as
\begin{eqnarray}
    \label{V_H}
  V_{H}(X) \simeq 
  \frac{1}{2} \xi^2 m_{3/2}^2 \left| X \right|^2,
\end{eqnarray}
where $\xi^2$ is a dimension-less 
parameter of the order one, which we assume to be 
positive. 
We have neglected higher order terms 
because they are suppressed by the gravitational scale.

On the other hand, effects of SUSY breaking through the
gauge-mediation mechanism also generate a potential $V_G$ to the PQ
$X$ field.  In most models of the gauge-mediation\cite{DNetal,IY} SUSY
is broken by some non-perturbative dynamics in a ``{\it hidden}''
sector and its effects are fed down to a messenger sector.  In the
messenger sector, a gauge-singlet chiral multiplet $Z$ is supposed to have a
$A$-component vacuum expectation value (vev) $\left< Z \right>$ and a
$F$-component vev $\left< F_Z \right>$ as well.  This singlet couples
to messenger quark multiplets $q_M$ and $\overline{q}_M$
which are $\bf3$ and $\bf3^\ast$ in SU(3)$_C$,%
\footnote{
In fact, we also introduce messenger leptons 
so that non-color SUSY particles acquire soft masses.
}
in the superpotential
\begin{eqnarray}
    W_M = \lambda_M ~Z~ q_M \overline{q}_M
    \label{sp-M}
\end{eqnarray}
with $\lambda_M$ a dimension-less coupling.
The superpotential (\ref{sp-M}) induces
a mass of the messenger quark $\Psi_q$ as
$m_{\Psi_q} = M$ 
and the messenger squarks $\widetilde{q}_\pm$
obtain masses squared as
$m_{\widetilde{q}_\pm}^2 = M ( M \pm \Lambda )$.
Here we define $M$ and $\Lambda$ by
\begin{eqnarray}
    \lambda_M \left< Z \right> &\equiv& M,
\\    
    \lambda_M \left< F_Z \right> &\equiv& M \Lambda.
\end{eqnarray}
Then the SUSY is broken in the messenger sector 
and its effects are mediated to the ordinary sector
through the SM gauge interaction by integrating out 
the heavy messenger fields.

The PQ squarks $\widetilde{Q_1}$ and $\widetilde{Q_2}$,
similar to the squarks in the standard model sector,
obtain soft SUSY breaking masses 
from two-loop diagrams
of the messenger multiplets, gluon and gluino.
Then the SUSY breaking effects are transmitted to 
the PQ $X$ field by  ``one-loop'' diagrams of 
$\widetilde{Q_1}$ and $\widetilde{Q_2}$.

We made an explicit calculation of the effective potential
for the PQ scalar $X$, $V_G(X)$, along the flat direction
induced by the above gauge-mediation mechanism.
Here we present only the final result and
the details are found in Appendix.\footnote{
After submitting the paper, we were informed that the saxion potential
from the gauge-mediated SUSY breaking effect was also calculated in 
ref.\cite{AGLR} in a 
different manner. See section 5.2 of \cite{AGLR} for more detail. Our result
is in agreement with that of \cite{AGLR}.}
To estimate the minimum of the potential,
the derivative of the potential with respect to $X^2$
is sufficient rather than $V_G$ itself.
The result is given by\footnote{$X$ here should be understood as its modulus
$|X|$. We use this notation to avoid unnecessary complications.}
\begin{eqnarray}
  \frac{ \partial V_G(X^2) }{\partial X^2} 
  &=& ~-~ \frac{ 8 \alpha_{s}^2 }{ ( 4 \pi )^4 }
        \frac{ M^2 \Lambda^2 }{ X^2 }
        \log^2 \left( \frac{ \lambda_{P}^2 X^2 }{M^2} \right).
      \label{V_G}
\end{eqnarray}
Here we have presented the asymptotic form which is valid only for $X
\gg M$. This is because for $X \lsim M$, the gauge-mediation generates
a sizable negative curvature for the $X$ field in a standard manner,
which shifts the vev of the field far away from the origin. Thus we
expect that the minimum of $X$ is much larger than $M$. This is a good
news for phenomenology because there is some chance that the resulting
PQ scale falls into the cosmological axion window, which is indeed the
case, as we will see shortly.

The potential along the flat direction $X$ generated by the two SUSY 
breaking effects is thus given by
\begin{eqnarray}
    V_{eff}(X) = V_H(X) + V_G(X).
\end{eqnarray}
Using eqs.~(\ref{V_H}) and (\ref{V_G})
we can estimate the vev of the PQ scalar field,
$X_c \equiv \left< X \right>  
\gg M$, by looking into the minimization condition of the potential
\begin{eqnarray}
    0 = \frac{ \partial V_H(X) }{ \partial X}
       + \frac{ \partial V_G(X) }{ \partial X},
\end{eqnarray}
which yields
\begin{eqnarray}
  X_{c} &\simeq&
  \frac{ 1 }{ \xi \pi }
  \left( \frac{ \alpha_s }{ 4 \pi } \right)
  \frac{ M \Lambda }{ m_{3/2} }
  \log \left( \frac{ \lambda_{P}^2 X_{c}^2 }{ M^2 } \right).
\end{eqnarray}

The PQ scalar field thus develops a vev, and 
the U(1)$_{PQ}$ symmetry is spontaneously broken and the PQ scale
is given as $f_{PQ} = X_c$.
We would like to emphasize that in the present axion model
the PQ scale is determined by only the SUSY breaking scales
without introducing any other mass scales.
We present the numerical values of the PQ scale in Fig.1.
If we take the parameters of the messenger sector as
$\Lambda$ = 40 TeV and $M$ = 50 TeV
the PQ scale is approximately given by 
\begin{eqnarray}
    f_{PQ} = X_c \simeq \frac{10^{11 }~\mbox{GeV}}{\xi}
    \left( \frac{1~\mbox{MeV}}{ m_{3/2} } \right),
\label{eq:PQscale}
\end{eqnarray}
for $\lambda_{P}$ = 1.  Therefore this model
provides a simple description of the PQ mechanism in the gauge-mediated
SUSY breaking theories. Note that in the simplest structure of the
messenger sector, our choice of $\Lambda$ yields the right-handed
slepton mass of about 90 GeV. On the other hand, the choice of $M$ is
somewhat arbitrary, though $M$ must be larger than $\Lambda$ to avoid a
negative mass squared for the messenger squark, and thus one should regard 
(\ref{eq:PQscale}) as the lower bound of the PQ scale for a given gravitino 
mass. One should keep this point in mind, though in the following we take
$M=50$ TeV for our representative value. 

In SUSY theories, the axion forms a chiral multiplet. In addition to
the axion, which can be chosen as the imaginary part of the complex
scalar $X$, the multiplet contains a saxion $s$, the real part of $X$,
and an axino $\tilde{a}$, the fermionic component of the multiplet.
Parameterizing $X$ = $ ( s/\sqrt{2} + X_c) \exp \left( i a/(\sqrt{2}
  X_c) \right)$ the mass of the saxion is easily evaluated as
\begin{eqnarray}
  m_{s} &=& \xi m_{3/2}.
\end{eqnarray}
Thus the saxion acquires a mass comparable to the gravitino, while the axion 
$a$ gets a mass from the QCD effect as $m_a \simeq f_\pi m_\pi/f_{PQ}$.


We shall next consider cosmology of this axion model to see whether
or not it is cosmologically viable.  The axion relic abundance from 
the misalignment is given as \cite{Turner,KolbTurner}
\begin{equation}
\Omega_a h^2 \sim 0.2 
\theta^2 \left(\frac{f_{PQ}}{10^{12} \mbox{GeV}}\right)^{1.18}
\label{eq:axion-abundance}
\end{equation}
with $\theta$ being the initial misalignment angle ($|\theta| <\pi$) and $h$ 
 the Hubble constant in units of 100 km/sec./Mpc. In
ignorance of the misalignment angle and with theoretical uncertainty
of order unity in the evaluation of Eq.~(\ref{eq:axion-abundance}), we
could say that the axion may be able to constitute the dominant
component of the dark matter of the universe if
\begin{equation}
   10^{11} \mbox{GeV} \lsim f_{PQ} \lsim 10^{13} \mbox{GeV}.
\end{equation}
The same argument implies an upper bound of the axion scale $ f_{PQ}
\lsim 10^{13} \mbox{GeV}$.  On the other hand, the cooling of the SN 1987A
 puts a lower bound of about 
$10^{9}$ GeV \cite{cooling}. Thus the following range 
\begin{equation}
  10^{9} \mbox{GeV} \lsim f_{PQ} \lsim 10^{13} \mbox{GeV}
\label{eq:f_PQ:allowed}
\end{equation}
is regarded as the allowed region of the PQ scale. With
eq. (\ref{eq:PQscale}), this in turn identifies the allowed range 
of the gravitino mass in this scenario:
\begin{equation}
     100 \mbox{MeV} \gsim \xi m_{3/2} \gsim  10 
\mbox{keV}. \label{eq:gravitino-mass-range}
\end{equation} 
Remarkably the gauge-mediated SUSY breaking naturally gives the
gravitino mass in this range.  The region of $m_{3/2}$ is somewhat
shifted if we take a different value of $M$, but still the gravitino
mass range fits with the gauge-mediation mechanism as far as 
$M \lsim 10^9$ GeV with $\Lambda \simeq 40$ GeV fixed. 
On the other hand, this model
does not give the possible hadronic axion window $f_{PQ} \sim 10^6$
GeV.

Let us next consider cosmology of the saxion.  It dominantly decays to two 
axions through the interaction given by
\begin{equation}
  \partial_{\mu}X^{\dagger} \partial^{\mu} X 
 =\frac{1}{2} \partial_{\mu} s \partial^{\mu} s
 +\frac{1}{2} \partial_{\mu} a \partial^{\mu} a
 +\frac{1}{\sqrt{2}{X_c}} s  \partial_{\mu} a \partial^{\mu} a +\cdots.
\end{equation}
The decay width is estimated as
\begin{equation}
   \Gamma (s \rightarrow aa)
  = \frac{1}{64 \pi} \frac{m_s^3}{f_{PQ}^2},
\end{equation}
from which we find the lifetime
\begin{equation}
  \tau_s 
\simeq 1.3 \times 10^{9} \mbox{sec}
\left( \frac{1 \mbox{MeV}}{m_{s}} \right)^5
\simeq \frac{1.3 \times 10^{9} \mbox{sec}}{\xi^5}
         \left( \frac{1 \mbox{MeV}}{m_{3/2}} \right)^5.
\end{equation}
It varies widely from $10^{-1}$--$10^{19}$ sec. for the gravitino mass range
(\ref{eq:gravitino-mass-range}).  
With this long lifetime, we have to see its cosmological consequences
carefully. 

To go further, we need to estimate the saxion abundance. The
saxion decouples from the thermal bath at the temperature
about \cite{RTW}
\begin{equation}
  T_{dec} \simeq 10^{9} \mbox{GeV} 
     \left(\frac{f_{PQ}}{10^{11} \mbox{GeV}}\right)^2.
\end{equation}
If the decoupling temperature were lower than the reheat temperature
after the inflation, then the yield of the saxion (defined by the
number density of the saxion divided by the entropy density) would be
$Y_s \sim 10^{-3}$. However the gravitino problem in the GMSB suggests 
relatively low $T_R$. For the gravitino mass range considered here, the closure
limit of the gravitino gives \cite{MMY,GMM}
\begin{equation}
 T_R \lsim 10^6 \mbox{GeV} 
  \left(\frac{m_{\tilde g}}{300\mbox{GeV}}\right)^{-2}
  \left(\frac{m_{3/2}}{1 \mbox{MeV}}\right),
\label{eq:TR-gravitino}
\end{equation}
where $m_{\tilde g}$ denotes the gluino mass.  So the decoupling
temperature is higher than the reheat temperature, and saxions are not
thermalized.\footnote{
When $f_{PQ}$ is closed to its lower bound
$10^9$ GeV the decoupling temperature can be larger than the reheat
temperature. Subsequent argument in the text does not apply for this
case. For example, the abundance of the saxion simply  becomes $m_sY_s \sim
10^{-4}$ GeV $(m_s/100$ MeV) instead of (\ref{eq:saxion-abundance-thermal}).
  We have considered cosmology of the saxion in
this exceptional case separately and checked that this survives cosmological 
constraints described below.} 
However saxions are produced by scattering {\em e.g.} $q
\bar q \rightarrow s g$ in the thermal bath, with the yield
\cite{KimKim,ChangKim}
\begin{equation}
Y_s \sim 10^{-3} \left(\frac{T_R}{T_{dec}}\right)
\sim 10^{-6} \left(\frac{f_{PQ}}{10^{11} \mbox{GeV}}\right)^{-2}
      \left(\frac{T_R}{10^6 \mbox{GeV}}\right),
\label{eq:saxion-yield} 
\end{equation}
and thus
\begin{equation}
  m_s Y_s \sim 10^{-9} \mbox{GeV} \left(\frac{m_s}{1 \mbox{MeV}} \right)
         \left(\frac{f_{PQ}}{10^{11} \mbox{GeV}}\right)^{-2}
       \left(\frac{T_R}{10^6 \mbox{GeV}}\right).
\label{eq:saxion-abundance-thermal}
\end{equation}
     
The saxion is also produced as its coherent oscillation.  An
inspection shows that with the low reheat temperature given by
(\ref{eq:TR-gravitino}) the saxion starts to oscillate while the
inflaton obeys coherent oscillation (before the reheating process is
completed). We estimate the abundance to be
\begin{equation}
   m_s Y_s \sim \frac{1}{8} \left(\frac{s_i}{M_G}\right)^2 T_R
     \simeq 10^{-10} \mbox{GeV}
         \left(\frac{s_i}{f_{PQ}}\right)^2    
     \left(\frac{f_{PQ}}{10^{11}\mbox{GeV}}\right)^2 
     \left(\frac{T_R}{10^{6}\mbox{GeV}}\right),
\label{eq:saxion-abundance-osc}
\end{equation}
where $s_i$ is the initial amplitude of the saxion oscillation,
which is expected to be of order $f_{PQ}$.

In the above, we have assumed that the vacuum expectation value of the
field $X$ is displaced from the origin during the reheating process:
otherwise thermal effects would generate a positive mass for
$X$, trapping the $X$ field at the origin until the universe cools down
below the weak scale. In this case, the above estimate is no longer
valid, and we expect that the relic abundance of the saxion will be
much larger, which may be problematic. The displacement of the field
$X$ may be achieved by fluctuations of the field during the de Sitter
expansion in the inflationary phase.  In the following, we use the
maximum of (\ref{eq:saxion-abundance-thermal}) and
(\ref{eq:saxion-abundance-osc}) as the saxion abundance. It remains
constant as time goes on as far as no entropy production occurs. Note
that the abundance is proportional to the reheat temperature.

Let us next consider bounds on the abundance of the saxions and the
axions produced by the saxion decay. Existence of exotic particles
around the cosmic temperature $T\sim 1$ MeV may accelerate the
expansion of the universe and increase the neutron-to-proton number
ratio at the decoupling, resulting in too much $^4$He
abundance. Roughly speaking, the abundance of such  exotic particles
should not exceed that of one neutrino species, namely
\begin{equation}
   m_sY_s \lsim 10^{-4} \mbox{GeV}.
\end{equation}
This is not  significant.

The energy
densities of the saxions and the produced axions may alter the evolution
of the universe at much later time, {\it e.g.} the age of the universe
and the time of
matter-radiation equality.  However, if the abundance of the produced saxions
 does not exceed the critical density of the universe 
(divided by the present entropy density), namely
\begin{equation}
    m_s Y_s \lsim 3.6 \times 10^{-9} h^2 \mbox{GeV},
\label{eq:saxion-abundance-safe}
\end{equation}
then the standard cosmology is not affected. We find that the
condition (\ref{eq:saxion-abundance-safe}) is easily satisfied for a
certain range of the reheat temperature.

The next bounds we consider are those from rare decay modes of the saxion.
Since the saxion is light, it does not decay to gluons. On the other
hand, it generally decays to photons. The branching ratio of the
radiative decay of the saxion, $B$, is typically of the order
$(\alpha/4\pi)^2\sim 10^{-6}$. For $\tau_s \lsim 10^{6}$ sec. the 
constraint on the radiative decay comes from the
photodissociation of the light elements (for a recent analysis, see
ref. \cite{HKKM}).  For $\tau_s \gsim 10^6$ sec., the non-observation of the
distortion of the cosmic microwave background gives a tighter bound on
the abundance of the produced photons: $B \times m_sY_s \lsim 2 \times
10^{-13}$ GeV $(10^{10}$sec.$/\tau_{s})^{1/2}$ for $\tau_s \sim
10^{6}$--$10^{10}$ sec. and $B \times m_sY_s \lsim 2 \times 10^{-14}$
GeV $(10^{12}$sec.$/\tau_s)^{2/3} (\Omega_0 h^2)^{-1/3}$ for $\tau_s \sim
10^{10}$--$10^{12}$ sec. \cite{Fixsenetal}.  Comparing the above
bounds with the predicted abundance
(\ref{eq:saxion-abundance-thermal}) or (\ref{eq:saxion-abundance-osc})
multiplied by the very small branching ratio $B$, we find that the
saxion abundance easily survives these bounds even with relatively a
high reheat temperature close to the upper bound allowed by the
gravitino problem.  On the other hand, when the saxion decays after the
recombination ($\tau_s \gsim 10^{12}$ GeV), a stringent constraint comes
from the diffuse X-ray backgrounds. The strongest one is $B \times m_s Y_s
\lsim 10^{-17}$ GeV for $\tau_s \sim 10^{17}$
sec. \cite{KawasakiYanagida}.  We find that this severe constraint is
fulfilled if the reheat temperature is lower than 10 GeV so that the
saxion abundance is highly suppressed.

Summarizing the above arguments, we conclude that
our model can be viable in view of the saxion cosmology.

It is noteworthy to mention that
the case where the saxion abundance exceeds (\ref{eq:saxion-abundance-safe})
is not immediately ruled out. Rather 
the saxion may modify the universe's evolution in an interesting
way. Probably the most interesting case \cite{CKK,KimKim} is that the
saxion energy density once dominates the universe before it decays to
axions. Then the produced axions behave as an extra contribution to
the radiation when we consider the cosmological developments, which
delay the epoch of the matter-radiation equality.  In this case, the
saxion may play a role of the late decaying particle which will
reconcile the standard cold dark matter (CDM) dominated universe for
$h > 0.5$ with the large scale structure formation, if the following
is fulfilled \cite{KimKim}
\begin{equation}
   \left(\frac{\tau_s}{1 \mbox{sec}}\right) 
  \left(\frac{m_sY_s}{1 \mbox{MeV}}\right)^2 \simeq 0.55 ((h/0.2)^2-1)^{3/2}.
\end{equation}
 In particular, our
model  provides a concrete realization of 
ref.~\cite{ChangKim}, where a hypothetically light saxion is proposed 
to play this role. 

Finally we will look into the axino, the fermionic superpartner of the
axion. Radiative corrections with the messenger fields give an axino mass.
In our case,
the contribution appears at the three-loop level, and it is $\sim
m_{3/2}^2/\Lambda \ll m_{3/2}$. The axino mass is also induced by
supergravity effects and in general it becomes of the order of the
gravitino mass \cite{GotoYamaguchi,CKN,ChunLukas}.  An inspection
shows that, however, in this simple model, the leading contribution of
the order $m_{3/2}$ is cancelled, and the non-vanishing contribution
is $\sim (f_{PQ}^2/M_G^2) m_{3/2} \ll m_{3/2}$.  Thus the axino is
much lighter than the gravitino. The axinos are produced by scattering
after the inflationary epoch and we expect that the yield of the
axinos is more or less the same as that of the saxions
(\ref{eq:saxion-yield}). In our case, since the axino is much lighter
than the saxion, the axino energy density is always smaller than the
saxion's one and thus it is harmless.  Another possible worry is that the
gravitino decay to an axino and an axion is kinematically
allowed. However, with the mass range of the gravitino considered
here, the lifetime is much longer than the age of the universe, hence
the gravitino is essentially stable. Thus the light axino appeared in
this model does not alter the argument above and it is cosmologically
harmless.

%
To conclude, we proposed the hadronic axion model in the gauge-mediated SUSY
breaking theories.  The model gives a simple description on the PQ
symmetry breaking mechanism which is solely governed by the SUSY
breaking physics.  It should be noted that the PQ scale 
is inversely scaled to the gravitino
mass and the resulting PQ scale naturally falls into the axion window,
for the gravitino mass range favored by the gauge-mediation.  Moreover
the axion becomes a dark matter candidate for $m_{3/2} \sim 10$ keV--1
MeV.  Next we investigated cosmological implications of the other
particles in the axion supermultiplet, namely the saxion and the
axino, and showed that our model is cosmologically acceptable with
sufficiently low reheat temperature which is suggested by the
gravitino problem. Finally we pointed out that the saxion in this
model may play a role of the late-decaying particle, causing the
delay of the matter-radiation equality, and consequently reconciling
the standard CDM with the structure formations of the universe.  Details of
this issue will be discussed elsewhere \cite{AY}.

\section*{Acknowledgments}
TA would like to thank M. Kawasaki for helpful discussions.
The work of MY was  supported in part by 
the Grant--in--Aid for Scientific Research from the Ministry of 
Education, Science and Culture of Japan No.\ 09640333.

\clearpage
\appendix
\section*{Effective Potential for PQ Scalar Field $X$}
%
Here we derive the potential for the PQ scalar field $X$ induced by
the SUSY breaking effects through gauge-mediation mechanism.  We
calculate an effective potential for $X_c$, the classical value of the
$X$ field.  As the ordinary squarks, the PQ squarks
receives the SUSY breaking effects from the messenger sector by the QCD
interaction at two-loop level.  Then the effects are transmitted to
$X$ by the Yukawa interaction (\ref{sp-PQ}) at three-loop order.  The
Feynmann diagrams for the vacuum energy which should be estimated are
shown in Figs. 2, 4 and 5.  In our calculation we expand it in terms
of the $M \Lambda$ and estimate the leading term of ${\cal O}((M
\Lambda)^2)$, i.e., the leading term with $F$-vev of the  messenger
multiplet.  From the Yukawa interaction (\ref{sp-PQ}) 
the PQ quark and squark acquire a mass $M_Q = \lambda_P X_c$.

First, we estimate the vacuum energy induced by 
the D-term scalar potential which diagram is shown 
in Fig. 2.
This diagram gives
\begin{eqnarray}
   i V_D &=& 
       2 \times 2 g_s^4
       \int \frac{ d^4 k }{ (2 \pi )^4 }
       \int \frac{ d^4 p }{ (2 \pi )^4 }
       \int \frac{ d^4 q }{ (2 \pi )^4 }
\nonumber \\
    &&\times
       \left[ 
           \frac{ 1 }{ [ k^2 - m_{\widetilde{q}_+}^2 ]
                       [ (k+p)^2 - m_{\widetilde{q}_-}^2 ]  }
       \right]
       \left[ 
           \frac{ 1 }{ [ q^2 - M_{Q}^2 ]
                       [ (q+p)^2 - M_{Q}^2 ]  }
       \right],
\end{eqnarray}
where the messenger squark masses are $m_{\widetilde{q}_\pm}$
=$ M(M \pm \Lambda)$.
Then we expand the RHS in terms of $(M \Lambda )$ as
\begin{eqnarray}
    i V_D &=& 
       4 g_s^4
       \int \frac{ d^4 k }{ (2 \pi )^4 }
       \int \frac{ d^4 p }{ (2 \pi )^4 }
       \int \frac{ d^4 q }{ (2 \pi )^4 }
       \int_0^1 dx
       \int_0^1 dy
      \frac{ 1 }{ \left[ q^2 + y (1-y) p^2 - M_Q^2 \right]^2 }
\nonumber \\
    &&\times
             \left[ 
           \frac{ 1 }{ \left[ k^2 + x (1-x) p^2 - M^2 \right]^2 }
           +
           \frac{ 3 ( 1-2x)^2 ( M \Lambda )^2 }
                { \left[ k^2 + x (1-x) p^2 - M^2 \right]^4 }
           + \cdots
       \right],
\end{eqnarray}
where the ${\cal O }( M \Lambda )$ term vanishes on making the integration 
on $x$.
Since what we are interested in are the SUSY breaking effects at the first
non-vanishing order,  
we pick up only the ${\cal O }( (M \Lambda)^2 )$ term.
Thus the effective potential from the D-term interaction 
is estimated as
\begin{eqnarray}
    i V_D &=& 
     g_s^4
       \int \frac{ d^4 k }{ (2 \pi )^4 }
       \int \frac{ d^4 p }{ (2 \pi )^4 }
       \int \frac{ d^4 q }{ (2 \pi )^4 }
       \int_0^1 dx
       \int_0^1 dy
\nonumber \\
    &&\times
           \frac{ 12 ( 1-2x)^2 ( M \Lambda )^2 }
                { \left[ q^2 + y (1-y) p^2 - M_Q^2 \right]^2
                    \left[ k^2 + x (1-x) p^2 - M^2 \right]^4 }.
    \label{veff-d}                    
\end{eqnarray}

Next, we turn to the vacuum energy with the gluon fields.
We first calculate the one-loop correction 
to the vacuum polarization from the messenger multiplets,
(see Fig.3.) and then 
estimate the vacuum energy shown in Fig.4.
Up to the ${\cal O }( (M \Lambda)^4 )$,
the result is given by
\begin{eqnarray}
    i V_g &=& 
       g_s^4
       \int \frac{ d^4 k }{ (2 \pi )^4 }
       \int \frac{ d^4 p }{ (2 \pi )^4 }
       \int \frac{ d^4 q }{ (2 \pi )^4 }
       \int_0^1 dx
       \int_0^1 dy
\nonumber \\
    &&\times
           \frac{ 36 ( 1-2x)^2 ( M \Lambda )^2 }
                { \left[ q^2 + y (1-y) p^2 - M_Q^2 \right]^2
                    \left[ k^2 + x (1-x) p^2 - M^2 \right]^4 }.
    \label{veff-gluon}                
\end{eqnarray}

Finally, the vacuum energy induced by the gluino (see Fig.5.)
is estimated as 
\begin{eqnarray}
    i V_{\widetilde{g}} &=&  - 
       g_s^4
       \int \frac{ d^4 k }{ (2 \pi )^4 }
       \int \frac{ d^4 p }{ (2 \pi )^4 }
       \int \frac{ d^4 q }{ (2 \pi )^4 }
       \int_0^1 dx
       \int_0^1 dy
\nonumber \\
    &&\times
           \frac{ 192 (1-y) ( 1-x)^3 ( M \Lambda )^2 }
                { \left[ q^2 + y (1-y) p^2 - M_Q^2 \right]^2
                    \left[ k^2 + x (1-x) p^2 - M^2 \right]^4 }.
  \label{veff-gluino}                    
\end{eqnarray}

Therefore summing up eqs. (\ref{veff-d}), (\ref{veff-gluon}) 
and (\ref{veff-gluino}), we obtain the effective potential
$V_G$ at the order $(M \Lambda)^2$ as
\begin{eqnarray}
    V_{G}( M_Q^2 ) &=&   
       i g_s^4
       \left( M \Lambda \right)^2
       \int \frac{ d^4 k }{ (2 \pi )^4 }
       \int \frac{ d^4 p }{ (2 \pi )^4 }
       \int \frac{ d^4 q }{ (2 \pi )^4 }
       \int_0^1 dx
       \int_0^1 dy
\nonumber \\
    &&\times
           \frac{ 48 x ( 1-x) }
                { \left[ q^2 + y (1-y) p^2 - M_Q^2 \right]^2
                  \left[ k^2 + x (1-x) p^2 - M^2 \right]^4 },
\end{eqnarray}
and by making a Wick rotation it leads to
\begin{eqnarray}
    V_{G}( M_Q^2 ) &=&   
       \frac{48 g_s^4}
           { (4 \pi )^6 }
       \left( M \Lambda \right)^2
       \int d k 
       \int d p
       \int d q 
       \int_0^1 dx
       \int_0^1 dy
\nonumber \\
    &&\times
           \frac{ x ( 1-x) kpq}
                { \left[ q + y (1-y) p + M_Q^2 \right]^2
                  \left[ k + x (1-x) p + M^2 \right]^4 }.
\end{eqnarray}
Instead of the effective potential itself, it is more convenient to calculate
its derivative. We find 
\begin{eqnarray}
    \frac{ \partial V_{G}( M_Q^2 ) }{ \partial M_Q^2 }
    &=&   
     -  \frac{ 96  g_s^4 }
           { (4 \pi )^6 }
        \left( M \Lambda \right)^2
       \int d k 
       \int d p
       \int d q 
       \int_0^1 dx
       \int_0^1 dy
\nonumber \\
    &&\times
           \frac{ x ( 1-x) kpq }
                { \left[ q + y (1-y) p + M_Q^2 \right]^3
                  \left[ k + x (1-x) p + M^2 \right]^4 }.
\end{eqnarray}
Now the momentum integrations are straight forward.
\begin{eqnarray}
    \frac{ \partial V_{G}( M_Q^2 ) }{ \partial M_Q^2 }
    &=&   
     -  \frac{ 8  g_s^2  }
           { (4 \pi )^6 }
       \frac{  \left( M \Lambda \right)^2 }{ M_Q^2 }
       \int_0^1 dx
       \int_0^1 dy
\nonumber \\
    &&\times
           \frac{ \frac{1}{z^2} - x(1-x) + x(1-x) \ln [z^2 x(1-x) ] }
                { \left[ \frac{ 1 }{z^2 } - x(1-x) \right]^2 },
\end{eqnarray}
where $z^2 = \frac{ M_Q^2 }{ M^2 y (1-y) }$.
We can perform the $x$ integration analytically.
(See ref.\cite{GMM}.)
Instead here we give an approximate  expression valid for $z^2 \gg 4 $ 
\begin{eqnarray}
        \frac{ \partial V_{G}( M_Q^2 ) }{ \partial M_Q^2 }
    &\simeq&   
     -  \frac{  8 g_s^4 }
           { (4 \pi )^6 }
       \frac{ \left( M \Lambda \right)^2 }{ M_Q^2 }
       \int_0^1 dy \ln^2 \left[ \frac{ M_Q^2 }{ M^2 y(1-y) } \right]
\nonumber
\\
   &\simeq& 
       - \frac{ 8  g_s^4 }
           { (4 \pi )^6 }
       \frac{ \left( M \Lambda \right)^2 }{ M_Q^2 }
       \ln^2 \left[ \frac{ M_Q^2 }{ M^2 } \right].
\end{eqnarray}
The PQ squark mass is expressed by the vev of $X$ 
and we obtain the final result 
\begin{eqnarray}
    \frac{ \partial V_{G}(X_c^2 ) }{ \partial X_c^2 }
    &\simeq& 
    -8 \frac{  g_s^4 }
           { (4 \pi )^6 }
         \frac{ \left( M \Lambda \right)^2 }{ X_c^2 }
       \ln^2 \left[ \frac{ \lambda_P^2 X_c^2 }{ M^2 } \right].     
\end{eqnarray}

\clearpage

\begin{figure}
    \label{fig-PQscale}
    \centerline{\epsfxsize=20cm \epsfbox{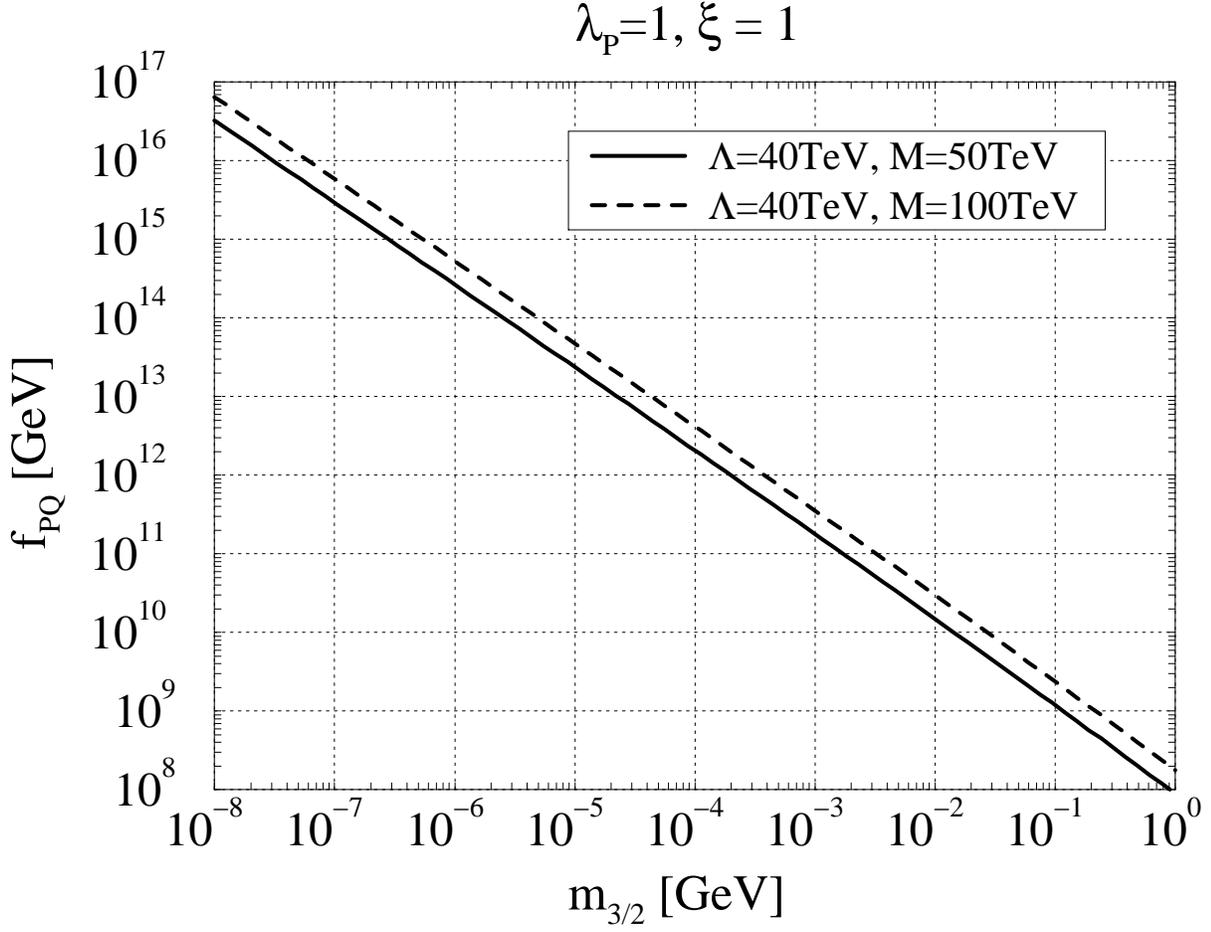}}
    \vspace{0.5cm}
\caption{
Values of the PQ breaking scale $f_{PQ}$.
Parameters for the messenger sector are listed in the figure.
}
\end{figure}
\clearpage

\begin{figure}
    \label{fig-fd-D}
    \centerline{\epsfysize=5cm \epsfbox{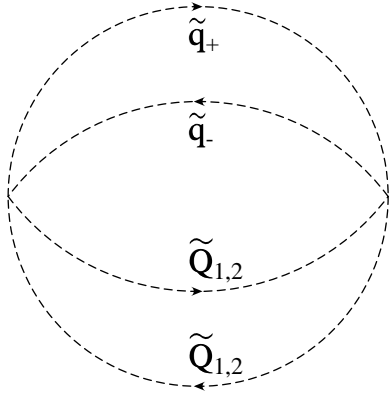}}
    \vspace{0.5cm}
\caption{
The vacuum diagram induced by the D-term scalar potential.
$\widetilde{q}_{\pm}$ are the messenger squarks and 
$\widetilde{Q}_{1,2}$ are the PQ squarks.}
\end{figure}
\vspace{3cm}

\begin{figure}
    \label{fig-fd-g2}
    \centerline{\epsfxsize=16cm \epsfbox{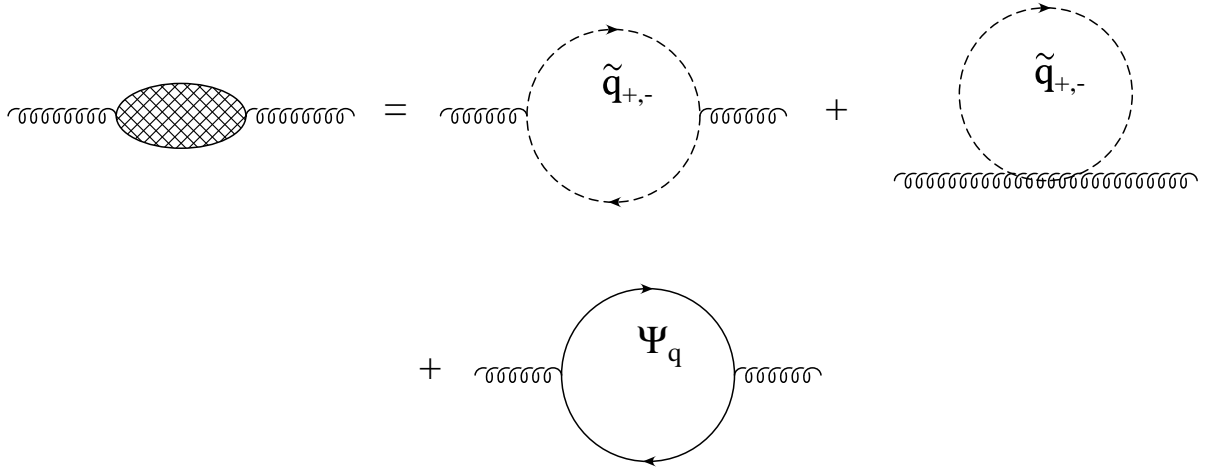}}
    \vspace{0.5cm}
\caption{
The one-loop diagrams of the vacuum polarization
induced by the messenger multiplets.
$\widetilde{q}_{\pm}$ are the messenger squarks and
$\Psi_{q}$ denotes the messenger quark.
}
\end{figure}

\clearpage

\begin{figure}
    \label{fig-fd-gluon}
    \centerline{(a){\epsfysize=5cm \epsfbox{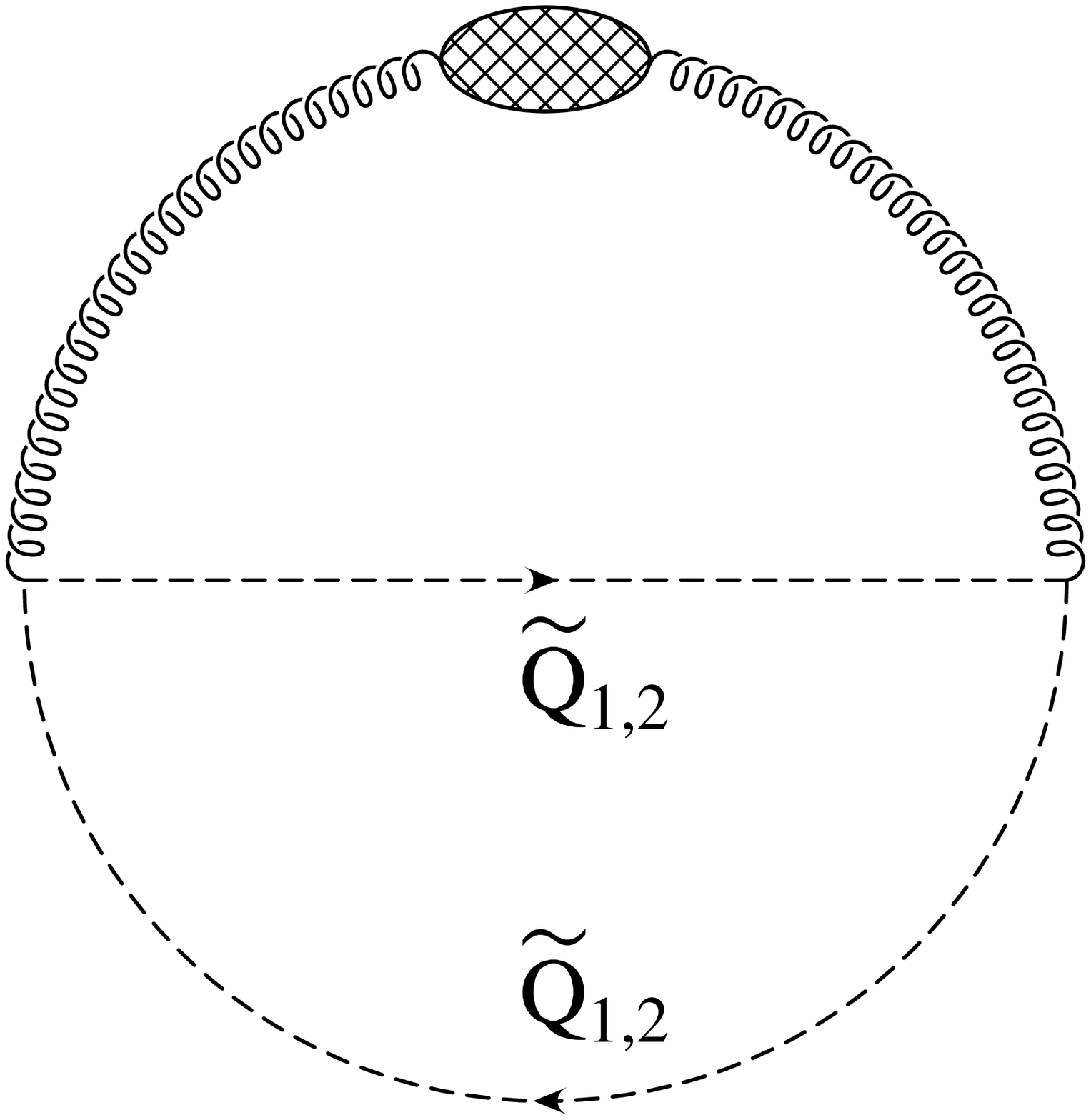}}\hspace{1cm}
                (b){\epsfysize=5cm \epsfbox{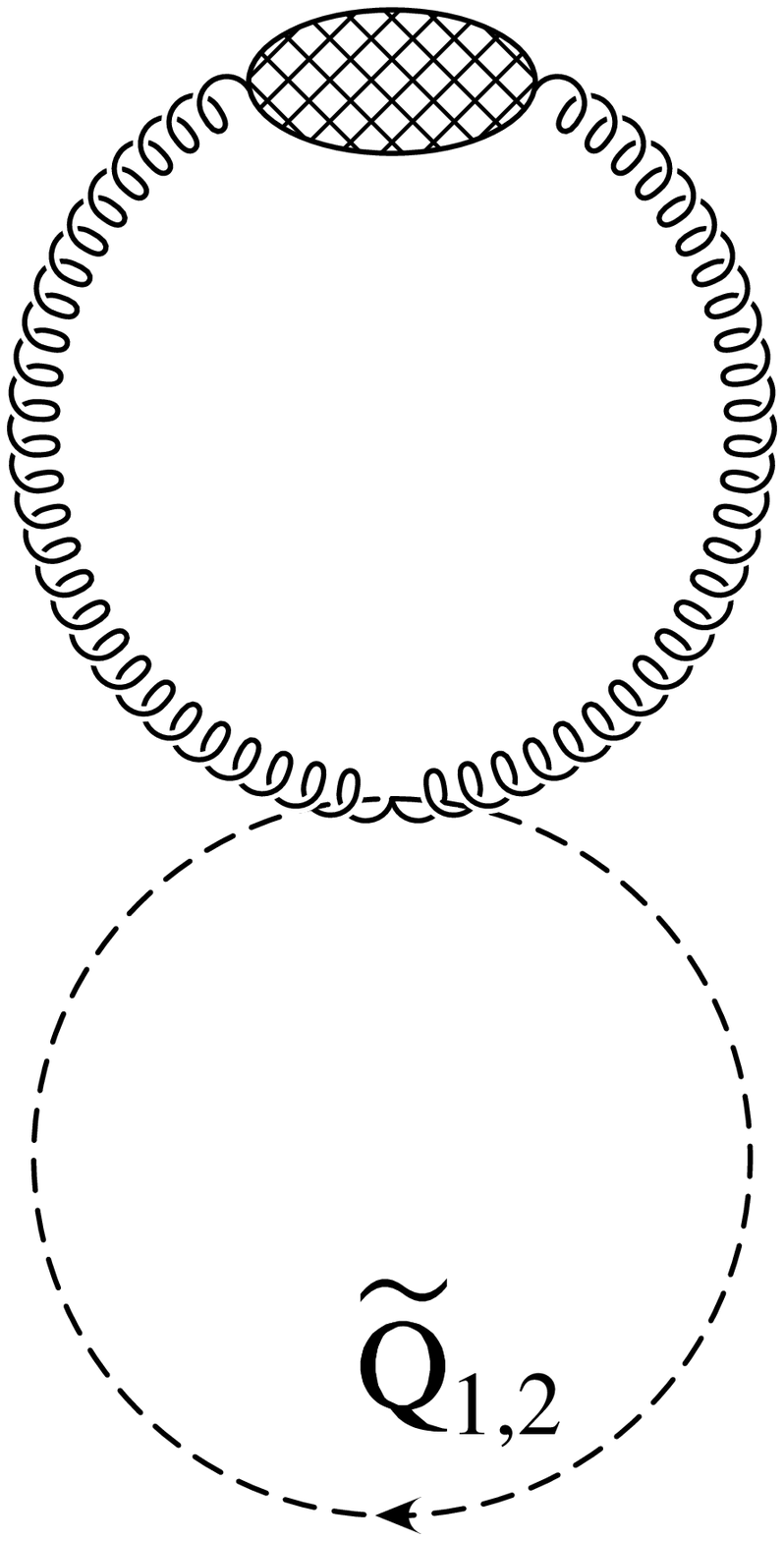}}\hspace{1cm}
                (c){\epsfysize=5cm \epsfbox{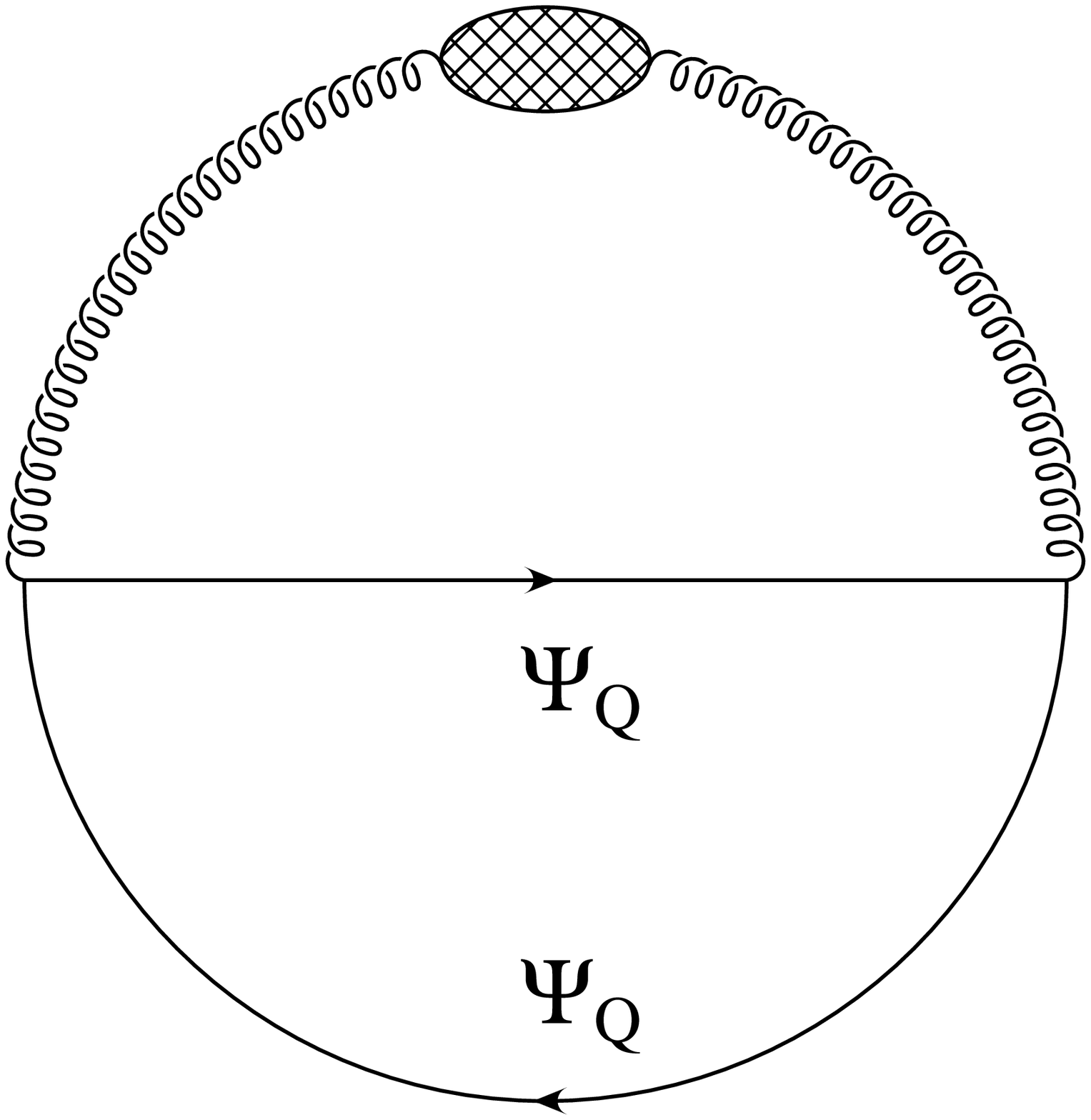}}}
    \vspace{0.5cm}
\caption{
The vacuum diagrams with gluon.
$\widetilde{Q}_{1,2}$ are the PQ squarks and
$\Psi_{Q}$ denotes the PQ quark.
A blob represents the one-loop vacuum polarization 
shown in Fig.3.
}
\end{figure}
\vspace{3cm}

\begin{figure}
    \label{fig-fd-gluino}
    \centerline{\epsfysize=6cm \epsfbox{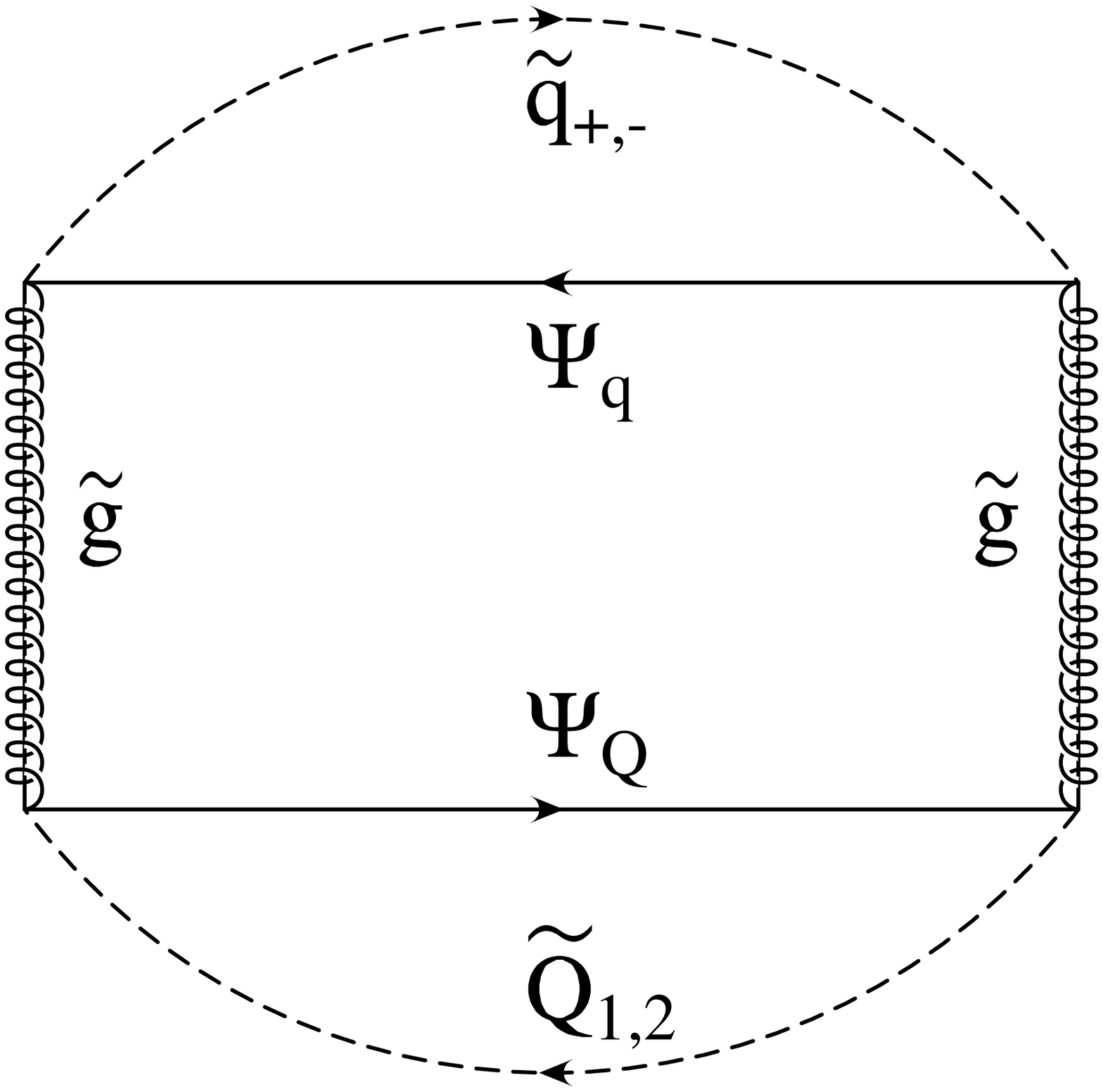}}
    \vspace{0.5cm}
\caption{
The vacuum diagram with gluino $\tilde g$.
}
\end{figure}


\begin{references}
\bibitem{PecceiQuinn}
R.D. Peccei and H.R. Quinn, Phys. Rev. Lett. {\bf 38} (1977) 1440;
 Phys. Rev. {\bf D16} (1977) 1791.

\bibitem{KSVZ}
J.E. Kim, Phys. Rev. Lett. {\bf 43} (1979) 103;
M.A. Shifman, A.I. Vainshtein and V.I. Zakharov,
Nucl. Phys. {\bf B166} (1980) 493.

\bibitem{KOY}
T. Kugo, I. Ojima and T. Yanagida, Phys. Lett. {\bf B135}  (1984) 402.

\bibitem{GMSB}
G.F. Giudice and R. Rattazzi,  hep-ph/9801271.
\bibitem{DNetal}
M. Dine and A.E. Nelson, Phys. Rev. {\bf D48} (1993) 1277; \\
M. Dine, A.E. Nelson and Y. Shirman, Phys. Rev. {\bf D51} 1362 (1995);\\
M. Dine, A.E. Nelson, Y. Nir and Y. Shirman, Phys. Rev. {\bf 53} 2658 (1996).


\bibitem{IY}
K.-I. Izawa and T. Yanagida, Prog. Theor. Phys. {\bf 95} (1996) 829; \\
K. Intriligator and S. Thomas,  Nucl. Phys. {\bf B473} (1996) 121.

\bibitem{AGLR}
N. Arkani-Hamed, G.F. Giudice, M.A. Luty and R. Rattazzi,
hep-ph/9803290.

\bibitem{Turner}
M.S. Turner, Phys. Rev. {\bf D33}  (1986) 889.

\bibitem{KolbTurner}
E.W. Kolb and M.S. Turner, {\sl The Early Universe},
Addison-Wesley (1990).  

\bibitem{cooling}
H.T. Janka, W. Keil, G. Raffelt and D. Seckel, Phys. Rev. Lett. 
{\bf 76} (1996) 2621.

\bibitem{RTW}
K. Rajagopal, M.S. Turner and F. Wilczek, Nucl. Phys. {\bf B358}  (1991) 447.

\bibitem{MMY}
T. Moroi, H. Murayama and M. Yamaguchi,
Phys. Lett. {\bf B303}  (1993) 289.
\bibitem{GMM}
A. de Gauvea, T. Moroi, and H. Murayama,
Phys. Rev. {\bf D56} (1997) 1281.

\bibitem{KimKim}
H.B. Kim and J.E. Kim,
Nucl. Phys. {\bf B433} (1995) 421.

\bibitem{ChangKim}
S. Chang and H.B. Kim, Phys. Rev. Lett. {\bf 77} (1996) 591.

\bibitem{HKKM} 
E. Holtmann, M. Kawasaki, K. Kohri and T. Moroi, hep-ph/9805405.

\bibitem{Fixsenetal}
D.J. Fixsen et al, Astrophys. J. {\bf 473} (1996) 473.

\bibitem{KawasakiYanagida}
M. Kawasaki and T. Yanagida, Phys. Lett. {\bf B399} (1997) 45.

\bibitem{CKK}
E.J. Chun, H.B. Kim and J.E. Kim, Phys. Rev. Lett. {\bf 72} (1994) 1956.

\bibitem{GotoYamaguchi}
T. Goto and M. Yamaguchi, Phys. Lett. {\bf B276} (1992) 103.

\bibitem{CKN}
E.J. Chun, J.E. Kim and H.P. Nilles, Phys. Lett. {\bf B287} (1992) 123.

\bibitem{ChunLukas}
E.J. Chun and A. Lukas, Phys. Lett. {\bf B357} (1995) 43.
\bibitem{AY}
T. Asaka and M. Yamaguchi, in preparation. 

%
\end{references}
\end{document}